
\input harvmac
\noblackbox
%
%

%

%

\def\Title#1#2{\ifx\answ\bigans \nopagenumbers
\abstractfont\hsize=\hstitle\rightline{#1}%
\vskip .5in\centerline{\titlefont #2}\abstractfont\vskip
.5in\pageno=0
\else \rightline{#1}
\vskip .8in\centerline{\titlefont #2}
\vskip .5in\pageno=1\fi}
\ifx\answ\bigans
 
scaled\magstep3
 
scaled\magstep3
 
scaled\magstep3
 
scaled\magstep3
 
scaled\magstep3
\else
 
scaled\magstep3
 
scaled\magstep3
 
scaled\magstep3
 
scaled\magstep3
 
scaled\magstep3
 
 \font\absi=cmmi10 scaled\magstep1
\font\absis=cmmi7 scaled\magstep1 \font\absiss=cmmi5 scaled\magstep1
\font\abssy=cmsy10 scaled\magstep1 \font\abssys=cmsy7 scaled\magstep1
\font\abssyss=cmsy5 scaled\magstep1 
scaled\magstep1
\skewchar\absi='177 \skewchar\absis='177 \skewchar\absiss='177
\skewchar\abssy='60 \skewchar\abssys='60 \skewchar\abssyss='60
\fi
%
%

\def\ajou#1&#2(#3){\ \sl#1\bf#2\rm(19#3)}

\def\frac#1#2{{#1 \over #2}}

\def\eps{\epsilon}

\def\mn{{\mu\nu}}

\def\Tr{{\rm Tr}}

\def\CL{{\cal L}}                       
\def\CG{{\cal G}}
\def\CM{{\cal M}}

\def\CJ{{\cal J}}
\def\p{\partial}

\def\ds{\raise.15ex\hbox{/}\kern-.57em\partial}
\def\Ds{\,\raise.15ex\hbox{/}\mkern-13.5mu D}
%

\lref\harvs{J. A. Harvey and A. Strominger,
Commun. Math. Phys. {\bf 151} (1993) 221.}
\lref\montolive{C. Montonen and D. Olive, Phys. Lett. {\bf 72B} (1977) 117.}
\lref\man{N. S. Manton,
Phys. Lett. {\bf 110B} (1982) 54.}
\lref\atiy{M. F. Atiyah and N. J. Hitchin,
Phys. Lett. {\bf 107A} (1985)
21. }
\lref\hitch{M. F. Atiyah and N.Hitchin, {\it The Geometry and Dynamics of
Magnetic Monopoles}, Princeton University Press (1988).}
\lref\lohe{M. Lohe, Phys. Lett. {\bf 70B} (1977) 325.}
\lref\osb{H. Osborn, Phys. Lett. {\bf 83B} (1979) 321.}
\lref\callias {C. Callias,
Commun. Math. Phys. {\bf 62} (1978) 213.}
\lref\birm{D. Birmingham, Phys. Rep. ......}
\lref\Bog{E. B. Bogomol'nyi, Sov. J. Phys. {\bf 24} (1977) 97.}
\lref\gary{G. W. Gibbons and N. Manton, Nucl. Phys. {\bf B274} (1986) 183.}
\lref\berndt{B. Schroers, Nucl. Phys. {\bf B367} (1991) 177.}
\lref\bernd{N. S. Manton and B. Schroers,
{\it Bundles over Moduli Spaces and the
Quantisation of BPS Monopoles}, DAMTP preprint DTP 93-05.}
\lref\ward{R. Ward, Phys. Lett. {\bf 158B} (1985) 424;
R. Leese, Nucl. Phys. {\bf B344} (1990) 33.}
\lref\atsing{Atiyah and Singer, M. F. Atiyah and I. M. Singer,
Proc. Nat. Acad. Sci. USA {\bf 81} (1984) 2597.}
\lref\jerome{J. P. Gauntlett, {\it Low-Energy Dynamics of Supersymmetric
Solitons}, to appear in Nucl. Phys. {\bf B}.}
\lref\divech{A. d'Adda, R. Horsley and P. di Vecchia, Phys. Lett.
{\bf 76B} (1978) 298.}
\lref\raj{ R. Rajaraman, {\it Solitons and Instantons}, North-Holland
1982.}
\lref\weet{E. Witten, Nucl. Phys. {\bf B202} (1982) 253.}
\lref\alvy{L. Alvarez-Gaum\'e and D.Z. Freedman, Commun. Math. Phys.
{\bf 80} (1981) 443.}
\lref\zumino{B. Zumino, Phys. Lett. {\bf 69B} (1977) 369.}
\lref\cole{S. Coleman, S. Parke, A. Neveu and C. M. Sommerfield,
Phys. Rev. {\bf D15} (1977) 544.}
\lref\witol{E. Witten and D. Olive, Phys. Lett. {\bf 78B} (1978) 97.}
\lref\ruback{P. Ruback, Nucl. Phys. {\bf B296} (1988) 669;
T. M. Samols, Commun. Math. Phys. {\bf 145} (1992) 149.}
\lref\eard{G. W. Gibbons and P. Ruback, Phys. Rev. Lett. {\bf 57} (1986) 1492;
R. C. Ferrel and D. M. Eardley, Phys. Rev. Lett. {\bf 59} (1987) 1617.}

\Title{\vbox{\baselineskip12pt
\hbox{EFI-93-09}
\hbox{hep-th/9305068}}}
{\vbox{\centerline{Low Energy Dynamics of N=2}
       \vskip2pt\centerline{Supersymmetric Monopoles}}}
\baselineskip=12pt
\bigskip
\centerline{Jerome P. Gauntlett}
\bigskip
\centerline{\sl Enrico Fermi Institute, University of Chicago}
\centerline{\sl 5640 Ellis Avenue, Chicago, IL 60637 }
\centerline{\it Internet: jerome@yukawa.uchicago.edu}
\medskip

\bigskip
\centerline{\bf Abstract}
It is argued that the low-energy dynamics of $k$ monopoles in
N=2 supersymmetric Yang-Mills theory
are determined by
an N=4 supersymmetric quantum mechanics
based on the moduli space of $k$ static monople solutions. This generalises
Manton's ``geodesic approximation" for studying the low-energy dynamics
of (bosonic) BPS monopoles. We discuss
some aspects of the quantisation and in particular argue
that dolbeault cohomology classes of the moduli space are related
to bound
states of the full quantum field theory.


\Date{5/93}

\newsec{Introduction}
A significant breakthrough in our understanding of the interactions
of monopoles followed the work of Manton \man.
He suggested
a method of approximating the low-energy interactions of
monopoles in Yang-Mills Higgs theory
in the BPS limit. In this limit there is a Bogomol'nyi bound \Bog\
on the static energy functional which implies the existence of
static multi-monopole solutions that saturate the bound.
The moduli space of static $k$-monopole
solutions, $\CM_k$,
is a finite dimensional subspace of the configuration space
that possesses a natural metric coming from the kinetic energy term in the
Lagrangian.
The low-energy dynamics of $k$
monopoles is
approximated by assuming that it is
geodesic motion on
$\CM_k$.
The metric on the moduli space for two monopoles
was constructed by Atiyah and Hitchin \atiy\ and has been used to develop
both the classical and quantum scattering problems (see \refs{\hitch
\gary-\berndt} and references therein).

Manton's approach can be generalised to study the low-energy
dynamics of ``solitons" of other systems including
Abelian Higgs vortices \ruback, ``lumps" of non-linear sigma models \ward\ and
extremal black holes \eard.
For this to work it
is crucial that the theory admits static mutli-soliton solutions.
Generically, this means that
Bogomol'nyi type bounds exist
and that the theory has a supersymmetric extension.

It is thus natural to extend these considerations
to study the low-energy interactions of
solitons in supersymmetric theories. Due to the presence
of fermions one is now neccesarily considering the low-energy
{\it quantum} theory. In a recent paper \jerome\ (see also \harvs),
we initiated this extension in the context of
the lump solutions of the
$N=2$ supersymmetric non-linear sigma model in $d=2+1$.
It was
shown that the low-energy dynamics of $k$ lumps
are described by an $N=2$ supersymmetric
quantum mechanics based on the moduli
space of $k$ static lump solutions. In the present work
we continue these investigations by studying the
dynamics of monopoles of $N=2$ supersymmetric Yang-Mills theory,
the $N=2$ extension of the BPS system\foot{In an interesting
recent paper \bernd\ the (non-supersymmetric)
quantum dynamics of iso-spinor fermions
coupled to the BPS system is studied.}.

We begin in section 2 with a review of BPS monopoles in the bosonic
theory.
The geometry of the moduli space of solutions is described and
some notation is introduced
that will be important in the later sections.
The low-energy dynamics of
monopoles in $N=2$ supersymmetric Yang-Mills
theory are discussed in section 3
and we show that they are determined by
an $N=4$ supersymmetric quantum mechanics (four real worldline supersymmetries)
based on the moduli space
of static BPS monopoles. In section 4 we discuss some aspects of the
quantum theory and section 5 concludes with some discussion.

\newsec{Review of BPS monopoles}
\subsec{Bogomol'nyi equations}
Let $A_m$ be an $SO(3)$ connection and $\Phi$ be a
Higgs field transforming in the
adjoint representation of $SO(3)$
governed by the Lagrangian density
\eqn\ymh{
\CL= -{1\over 4}\Tr F_{mn}F^{mn}+{1\over 2}\Tr
D_m\Phi D^m\Phi
}
where $F_{mn}$ is the curvature of the connection
and
$D_m=\p_m+[A_m,\ ]$ is the covariant
derivative.
Since there is no potential term for
the Higgs field, this being the BPS limit,
spontaneous symmetry
breaking of $SO(3)$ down to $U(1)$ is imposed by
demanding that
$\Tr\Phi^2=1$
as a boundary condition at infinity
\foot{Our choice of units is the same as in \gary.}.

It will be convenient to work in the $A_0=0$ gauge.
It is then necessary to impose
Gauss' Law, the $A_0$ equation of motion, as a constraint on
the time dependent physical fields:
\eqn\glawo{D_i\dot A_i + [\Phi,\dot\Phi]=0.}
In this gauge the Lagrangian is given by $L=T-V$ where the kinetic energy
$T$ is given by
\eqn\kin{T={1\over 2}\int d^3x\Tr(\dot A_i \dot A_i+\dot\Phi \dot\Phi)}
and the potential energy $V$ takes the form
\eqn\pot{V={1\over 2}\int d^3x\Tr(B_iB_i+D_i\Phi D_i\Phi),}
where $B_i={1\over 2}
\eps_{ijk}F_{jk}$ is the non-abelian magnetic field strength.
The total conserved energy is given by $E=T+V$.

In order to construct the static monopole solutions we need to minimise
the static energy functional $E=V$. It was shown by Bogomol'nyi \Bog\
that $V$ can be rewritten as
\eqn\bog{V=\int d^3x\Tr[{1\over 2}(B_i\mp
D_i\Phi)(B_i\mp D_i\Phi)]\pm 4\pi k,
}
where
\eqn\topc{k={1\over 4\pi}\int d^3x\p_i\Tr(B_i\Phi))}
is the monopole number which, if the fields are smooth,
is an integer topological invariant.
We thus deduce the Bogomol'nyi bound
\eqn\bogbd{V\ge 4\pi|k|.}
In each monopole class the static energy is minimised when the
bound is saturated which is equivalent to
the Bogomol'nyi equations
\eqn\boge{B_i=\pm D_i\Phi.}
The upper sign corresponds to positive $k$ or `` monopoles" and
the lower sign corresponds to negative $k$ or ``anti monopoles". From
now on we will restrict our considerations to monopoles, the extension
to anti monopoles being trivial. From \bogbd\ we see that the total energy
of $k$ static monopoles is $4\pi k$.
As is well known there are also electrically charged dyons in this model.
In the $A_0=0$ gauge there are no static dyon solutions; the dyons emerge
as time dependent solutions (see e.g. \gary\ for more details).

It will be useful in the following to
recall
the well known fact that the Bogomol'nyi equations are
equivalent to the
self-duality equations of pure Yang-Mills in $R^4$, restricted to be
translationally invariant in one direction \lohe.
Specifically, we define
a connection
$W_\mu$
on $R^4$ that is translationally invariant in the $x^4$
direction
via
\eqn\econ{W_i=A_i,\qquad W_4=\Phi
.}
If $G_{\mu\nu}$ is the field strength corresponding to  $W_\mu$,
then the self duality
equations for $W_\mu$,
\eqn\sd{
G_{\mu\nu}={1\over 2}\eps_{\mu\nu\rho\sigma}G_{\rho\sigma},}
are equivalent to the Bogomol'nyi equations \boge\ (upper sign).
Introducing the covariant derivative
on $R^4$,
$D_\mu=\p_\mu+[W_\mu,\ ]$,
Gauss' Law can be rephrased in terms of the
Euclidean connection as
\eqn\glawth{D_\mu\dot W_\mu=0.}
Finally an infinitesimal
gauge transformations on $(A_i,\Phi)$ can
be recast in the form
\eqn\gt{\delta W_\mu(x)=D_\mu\Lambda}
if the gauge parameter $\Lambda(x)$ is restricted to be
independent of $x^4$.

\subsec{Moduli space of BPS monopoles}
Denoting the space of finite energy field configurations
$\{W_\mu(x)\}$ by $\CA$ and the group of gauge transformations
by $\CG$, the configuration space of the BPS system is given by the
quotient $\CC=\CA/\CG$. That is, configurations related by
gauge transformations are identified. Letting $\dot W$ and $\dot V$
be two tangent vectors on $\CA$, a natural metric on $\CA$
is induced from that on $R^4$ via
\eqn\metric{\CG(\dot W,\dot V)=\int d^3x
\Tr (\dot W_\mu \dot V_\mu).}
A tangent vector to $\CC$ must satisfy Gauss' Law. From \glawth\ and
\gt\ it is clear that tangent vectors to $\CC$ are orthogonal to
the gauge orbits and hence that the metric $\CG$ is also well defined
on $\CC$. In fact \metric\ is just twice the kinetic energy functional
\kin. Since the potential energy \pot\ is gauge invariant, this confirms that
the configuration space for \ymh\ is indeed $\CC$.

The moduli space of $k$-monopoles,
$\CM_k\subset\CC$, is defined as the complete set of solutions to
the Bogomol'nyi equations
\boge\ within a given topological class $k$. A natural set of
co-ordinates for $\CM_k$ are the arbitrary parameters or moduli
$\{X^a,\ a=1,\dots,{\rm dim} (\CM_k)\}$ that determine the most general
gauge equivalence class of solutions $[W_\mu(x,X)]$.
In addition to Gauss' Law, tangent vectors to
$\CM_k$ must also obey the linearised Bogomol'nyi equations
\eqn\lbog{D_{[\mu}\dot W_{\nu]}=
{1\over 2}\eps_{\mn\rho\sigma}D_{\rho}\dot W_{\sigma}.}

There is a very close connection between zero modes in the fluctuations
about a particular monopole background and tangent vectors to $\CM_k$.
Using the co-ordinates $\{X^a\}$, we can express an arbitrary
tangent vector to $\CM_k$
as $\dot W_\mu=\dot X^a\delta_aW_\mu$ provided $\delta_aW_\mu$
satisfies
\eqn\zm{D_{[\mu}\delta_a W_{\nu]}=
{1\over 2}\eps_{\mn\rho\sigma}D_{\rho}\delta_a W_{\sigma}}
and
\eqn\glaw{D_\mu\delta_a W_\mu=0.}
Equation \zm\ is precisely the statement that $\delta_aW_\mu$ is a zero mode
and equation \glaw\ says that it is orthogonal to gauge modes.
A metric on $\CM_k$ is naturally
obtained from the restriction of the metric
$\CG$ \metric\ on $\CC$. In the co-ordinates $\{X^a\}$ this is given by
\eqn\metrict{\CG_{ab}=\int d^3x \Tr(\delta_a W_\mu\delta_b W_\mu).}
For the BPS system all zero modes are normalisable
(i.e.
finite $\CG_{ab}$) and there is a one to one correspondence between
normalisable zero modes and moduli. Note that this is not always true
(e.g. for the lumps of non-linear sigma models \ward).

Following closely the work in \harvs, we now
turn to a description of the zero modes that will be
important later.
Let $W_\mu(x,X)$ be a family of BPS monopole configurations
and $\eps_a(x,X)$ be a set of
arbitrary gauge transformations, then both
$\p_a W_\mu$
and $D_\mu \eps_a$ satisfy \zm, where $\p_a={\p\over \p X^a}$.
Zero modes about the configuration $W_\mu(x,X)$ may then be
constructed as the linear combination,
\eqn\zme{\delta_a W_\mu=\p_a W_\mu-D_\mu \eps_a,}
by demanding that the gauge parameters
are chosen to satisfy \glaw.
The gauge parameters $\eps_a$ can be viewed as
defining a natural
connection on $\CM_k$ with covariant derivative
\eqn\covd{s_a=\p_a+[\eps_a,\ ].}
The form of \zme\ then suggests that we should
consider the zero modes as the mixed curvature components
of a generalised connection $(W_\mu(x,X),\eps_a(x,X))$ defined on
$R^4\times\CM_k$. That is
\eqn\sat{[s_a,D_\mu]=\delta_a W_\mu.}
Using the definitions
\eqn\curvs{\eqalign{G_{\mn}&=[D_\mu,D_\nu]\cr
                  \phi_{ab}&=[s_a,s_b]\cr}}
Jacobi identities can be used to show that
\eqn\ids{s_aG_\mn=2D_{[\mu}\delta_aW_{\nu]}}
and
\eqn\jtwo{D_\mu\phi_{ab}=-2s_{[a}\delta_{b]}W_\mu.}
%
{}From \jtwo\ we further deduce that
the curvature $\phi_{ab}$ can be expressed
in the simple form
\eqn\curvt{\phi_{ab}=2(D_\mu D_\mu)^{-1}[\delta_a W_\nu,\delta_b W_\nu].}
The connection $(W_\mu,\eps_a)$
with curvature components
$(G_\mn,\delta_a W_\mu,\phi_{ab})$
is in fact the connection on
the ``universal bundle" introduced
by Atiyah and Singer \atsing.
The above identities can be used to
cast the Christoffel connection
associated to the metric \metrict\ in the form:
\eqn\chris{\Gamma_{abc}=\CG_{ad}{\Gamma^d}_{bc}=\int d^3x
\Tr(\delta_a W_\mu s_b\delta_c W_\mu)}

The space of field configurations $\CA$ inherits three almost
complex structures,
$\tilde J^{(m)}$, from those on $R^4$, $J^{(m)}$,via
\eqn\hk{[\tilde J^{(m)}\dot W]_\mu(x)=J^{(m)}_\mn \dot W_\nu.}
It is straightforward to show that $\tilde J^{(m)}$ satsify the
same algebra  as $J^{(m)}$ i.e.
$\tilde J^{(m)} \tilde J^{(n)}=-{\bf 1}\delta^{mn}+\eps^{mnp}\tilde J^{(p)}$.
Using the fact that each complex structure satisfies
\eqn\compli{\epsilon_{\mn\rho\sigma}=-(J^{(m)}_\mn J^{(m)}_{\rho\sigma}
+J^{(m)}_{\mu\sigma}J^{(m)}_{\nu\rho}+
J^{(m)}_{\mu\rho}J^{(m)}_{\sigma\nu})}
(no sum on $m$)
we can show that if $\dot W_\mu(x)$ is a tangent vector to $\CM_k$
(satisfies \glawth\ and \lbog) then so is
$J^{(m)}_{\mu\nu} \dot W_\nu$.
Thus the almost
complex structures on $\CA$, $\tilde J^{(m)}$, descend to give almost
complex structures
on $\CM_k$, $\CJ^{(m)}$. In the co-ordinates $\{X^a\}$ they take the form
\eqn\hkt{{{\CJ^{(m)}}_a}^b=\CG^{cb}\int d^3x {\CJ^{(m)}}_{\mu\nu} \Tr
(\delta_a W_\mu\delta_c W_\nu)}
To calculate the algebra satisfied by the $\CJ^{(m)}$
in this representation one needs to use the fact that the
zero modes $\delta_a W$ and the massive modes
\foot{Since the spectrum is continuous $i$ is a continuous parameter.}
$\psi_i$
in the fluctuations about a BPS
monopole configuration form a complete set of functions
\eqn\modes{\CG^{ab}\delta_a W_\mu^{\alpha}(x) \delta_b W_\nu^{\beta}(y)
+\sum_i \psi_{i\mu}^{\alpha}(x)\psi_{i\nu}^{\beta}(y)
=\delta^{\alpha\beta}\delta_{\mn}\delta^3(x-y)}
where $\alpha,\beta$ are $SO(3)$ indices.
This identity can also be used to show that
\eqn\hkid{{{\CJ^{(m)}}_a}^b\delta_b W_\mu=-J^{(m)}_{\mu\nu}\delta_a W_\nu.}
Using the formulae presented above we can also
show that the torsion vanishes
\eqn\tors{\p_{[a}{\CJ^{(m)}}_{bc]}=0.}
Although the discussion presented here is only schematic, it
can be rigorously shown that $M_k$ is indeed a hyperK\"ahler manifold \hitch.

\subsec{Geodesic approximation}
Let $W_\mu(x)$ correspond to a static multi-monopole
solution in the $A_0=0$ gauge. As we noted
the fluctuations about this solution contain massive
modes
in addition to the zero modes. To have a well defined
perturbation scheme, we need to introduce a collective co-ordinate
for each zero mode; these are the moduli
$X^a$. We then expand an arbitrary time dependent field
as a sum of massive modes with time dependent coefficients, by properly taking
into account the time dependence of the collective co-ordinates
(see, e.g. \raj).
A low-energy ansatz for the fields is obtained
by ignoring
the massive modes and demanding that the only time dependence is via the
collective co-ordinates. Thus we are led to the ansatz
\eqn\anso{\eqalign{W_\mu(x,t)&=W_\mu(x,X(t))\cr
A_0&=\dot X^a\eps_a\cr}}
After substituting this into the action we obtain an effective action
which is precisely that of a free particle propagating on the
moduli space with metric \metric \man. The equations of motion are
simply the geodesics on the moduli space.

The $A_0$ term in \anso\
is necessary to ensure that
the motion is constrained to be orthogonal to gauge transformations.
It can be justified more formally as follows.
Taking the dimension of
$X^a$ to be zero, the collective co-ordinate
expansion is an expansion in the number of time derivatives
$n=n_{\p_t}$.
Since the effective action is order
$n=2$
it is necessary to ensure that the equations of motion are
satisfied to order $n=0$ and $n=1$. To order $n=0$ this is true by assumption
that the collective co-ordinates are the moduli parameters. The order $n=1$
equation of motion is simply Gauss' Law
and we have explicitly solved this equation for $A_0$.
Since we are working in the $A_0=0$ gauge
we should really perform a gauge transformation on the
ansatz to stay in this gauge.
Of course this won't affect the result.
We have chosen to write the ansatz in this unorthodox way since
we construct a similar ansatz in the supersymmetric case.
\newsec{N=2 supersymmetric monopoles}
\subsec{Monopole solutions}
The N=2 supersymmetric
extension of the Yang-Mills Higgs model \ymh\ is described by
the Lagrangian density
\eqn\symt{
\eqalign{\CL=\Tr\{&-{1\over 4}(F_{mn})^2+{1\over 2}(D_m P)^2
+{1\over 2}(D_m S)^2 -{1\over 2}([S,P])^2\cr &+i\bar\chi\gamma^m
D_m\chi-\bar\chi\gamma_5[P,\chi]-i\bar\chi[S,\chi]\}
}}
where $A_m$ is an $SO(3)$ connection and the Higgs fields $P,S$ and the
Dirac fermions $\chi$ are in the adjoint representation of the $SO(3)$.
We choose a mostly minus metric and define $\gamma_5=i\gamma_0\gamma_1
\gamma_2
\gamma_3$.
The supersymmetry transformations are given by
\eqn\susyt{\eqalign{
\delta A_m&=i\bar\alpha\gamma_m\chi- i\bar\chi\gamma_m\alpha\cr
\delta P&=\bar\alpha\gamma_5\chi- \bar\chi\gamma_5\alpha,\qquad
\delta S=i\bar\alpha\chi- i\bar\chi\alpha\cr
\delta \chi&=\left(\sigma^{mn} F_{mn}-\Ds S +i\Ds P\gamma_5-i[P,S]\gamma_5
\right)\alpha\cr}
}
where $\alpha$ is a constant anticommuting Dirac spinor.
The Lagrangian can be obtained by the dimensional reduction of
supersymmetric Yang-Mills theory in six dimensions \divech.
The Lagrangian
is also invariant under chiral rotations which are a residuum of the
rotation group in the extra dimensions.

Since the potential for the Higgs fields has flat directions
corresponding to $[S,P]=0$,
the vacuum configuration is undetermined.
We again impose spontaneous
symmetry breaking by demanding
$\Tr S^2+\Tr P^2=1$
as a boundary condition at infinity. For $S$ and $P$ to commute in $SO(3)$
they must be proportional.
By performing a chiral rotation we can assume
that
only $S$ has
a vacuum expectation value.

By straightforward calculation one finds that the supersymmetry algebra
contains central charges \witol:
\eqn\susyalg{\{\bar Q,Q\}= 2\gamma_\mu P^\mu -2Q_E-2i\gamma_5Q_M}
where the electric and magnetic charges are
given by
\eqn\charges{Q_E=\int d^3x\p_i\Tr(E_i\Phi))\qquad Q_M=
\int d^3x\p_i\Tr(B_i\Phi))=4\pi k}
respectively. The electric charges can be understood as arising from
the components of the six dimensional momentum in the extra dimensions
whilst
the magnetic charges have a purely topological origin.
The presence of the central charges in the algebra implies that
there is a bound on the mass:
\eqn\bbb{M^2\ge Q_E^2 + Q_M^2.}
In the vacuum sector, the elementary particles form a supermultiplet
with charges and masses that saturate the
bound.

To discuss the soliton sectors we again work with the $A_0=0$ gauge.
In this gauge any static configuration has zero electric charge
and the bound \bbb\ is precisely the Bogomoln'yi bound we
discussed in the bosonic case \bogbd.
Thus, just as in the bosonic case, the bosonic fields are partitioned
into different topological sectors and for a given topological sector
the static energy is minimised by solving the
Bogomol'nyi equations
\eqn\susym{
\eqalign{{1\over 2} \epsilon_{ijk}F_{jk}&=D_i\Phi\cr
S&=\Phi\cr
P&=A_0=\chi=0\cr}}

These static
monopole solutions break half of the supersymmetries. By this we
mean that only half of the supersymmetry parameters in \susyt\ will
leave a solution of \susym\ invariant. Specifically, by introducing the
following projection matrix
\eqn\gfive{
\Gamma_5=-i\gamma_0\gamma_5,\qquad
\Gamma_5^2=1,\qquad\Gamma_5^\dagger=\Gamma_5}
and defining
\eqn\epsi{\alpha_\pm={1\pm\Gamma_5\over 2}\alpha}
one can show that $\alpha_+$ are the parameters of the
unbroken supersymmetry and $\alpha_-$ are the parameters of the
broken supersymmetry. This partial breaking of supersymmetry
can also be deduced directly from the supersymmetry algebra
\susyalg\ when the bound \bbb\ is saturated.

\subsec{Zero modes and supersymmetry}
We now turn to the construction of the zero modes in the
fluctuations about the monopole solution. The bosonic zero modes
are exactly the same as those for the  purely bosonic theory
and we discussed them in the last section. The fermionic zero modes
are time independent
c-number solutions to the Dirac equation in the presence of the
monopole:
\eqn\dirac{i\gamma^iD_i\chi-i[\Phi,\chi]=0.}
It is straightforward to show that the broken supersymmetry with
c-number parameters
generates two independent zero modes satisfying $\Gamma_5\chi=-\chi$.
These are the fermionic Goldstone modes corresponding to the broken
supersymmetry. In the one monopole sector these are the only
fermionic zero modes.

In the multi-monopole sectors there are other fermionic zero modes;
the Callias index theorem \callias\ states that in the $k$ monopole
sector there are $2k$ fermionic
zero modes. Following closely
an argument of Zumino's in the context of
instantons \zumino, we now show how the
bosonic and fermionic zero
modes form a supermultiplet with respect to the unbroken supersymmetry.
To exhibit this connection we first introduce some hermitean
euclidean
gamma matrices via
\eqn\egam{\Gamma_i=\gamma_0\gamma_i,\qquad\Gamma_4=\gamma_0}
satisfying
\eqn\ea{\{\Gamma_\mu,\Gamma_\nu\}=2\delta_{\mn}}
and we note that $\Gamma_5=\Gamma_1\Gamma_2\Gamma_3\Gamma_4$ justifying
the notation introduced in \gfive.
We next impose the following supersymmetric invariant restrictions
on the equations of motion
\eqn\res{\eqalign{\Gamma_5\chi&=-\chi\cr
A_0&=-P\cr
G_\mn&={1\over 2}\eps_{\mn\rho\sigma}G_{\rho\sigma}\cr}
}
and demand that all fields are time independent.

The equations of motion then read
\eqn\eom{\eqalign{G_\mn&={1\over 2}\eps_{\mn\rho\sigma}G_{\rho\sigma}\cr
\Gamma_\mu D_\mu\chi&=0\cr
D_\mu D_\mu P&= i[\chi^\dagger,\chi]\cr}}
and the unbroken supersymmetry transformations are given by
\eqn\unbs{\eqalign{
\delta W_\mu&=i\alpha_+^\dagger\Gamma_\mu\chi- i\chi^\dagger
\Gamma_\mu\alpha_+\cr
\delta \chi&=-2\Gamma_\mu D_\mu P
\alpha_+\cr
\delta P&=0\cr}
}
The equations of motion \eom\ are covariant with respect to these
transformations if $\alpha_+$ is a Grassmann odd spinor. If
$\alpha_+=\eps_+$ is a c-number spinor, the Dirac equation in
\eom\ is not covariant. However, if we impose the Dirac equation
then the other two equations are covariant. Thus
for each fermionic zero mode satisfying $\Gamma_5\chi=-\chi$,
we conclude that
\eqn\sbzm{\delta W_\mu=i\eps_+^\dagger\Gamma_\mu\chi-i\chi^\dagger\Gamma_\mu
\eps_+}
is a bosonic zero mode; it can be further checked that it
also satisfies \glaw.
This seems to imply there are four bosonic zero modes
for each fermionic zero mode, corresponding to the four real
independent parameters $\eps_+$. However, they are not all independent:
one can show that the four bosonic zero modes $(\delta_a W_\mu, J^{(m)}_\mn
\delta_a W_\nu)$ are paired with two fermionic zero modes $(\chi,C\chi^*)$
where $J^{(m)}_\mn$ are the three complex structures on $R^4$ and
the matrix $C$ satisfies $C^2=-1,\ \ C\Gamma_\mu=\Gamma_\mu^*C$.

Selecting a particular comlpex structure $J_\mn\equiv -J^{(3)}_\mn$ on
$R^4$ it will be convenient to pair
the bosonic and fermionic zero modes
via \harvs
\eqn\fzm{\chi_a = \delta_a W_\mu\Gamma^\mu\eps_+
}
where $\eps_+$ is a c-number spinor satisfying
\eqn\conds{\eqalign{\eps_+^\dagger\eps_+&=1\cr
J_{\mn}&=-i\eps_+^\dagger\Gamma_\mn\eps_+\cr
J_\mn\Gamma^\nu\eps_+&=i\Gamma^\mu\eps_+}}
(the last equation is derivable using a Fierz identity from the second).
Using \hkid\ we deduce that the fermionic zero modes satisfy
\eqn\fzmid{{\CJ_a}^b\chi_b=i\chi_a}
and hence that two bosonic zero modes
are paired with one fermionic zero mode.
The dimension of the moduli space of $k$ static
monopoles is $4k$ \hitch. From the above analysis
we conclude that there are $2k$
fermionic zero modes in the $k$ monopole sector,
in accord with the Callias index theorem \callias.

\subsec{Collective co-ordinate expansion}
To analyse the dynamics of the monopoles
we must introduce
a collective co-ordinate for each zero mode.
As we discussed in the last section, the bosonic collective
co-ordinates are co-ordinates on moduli space.
The fermionic collective
co-ordinates $\lambda^a$ are complex one
component Grassmann odd objects.
To construct a low-energy ansatz we ignore all non-zero modes
and use the supersymmetric pairing of the zero modes
\fzm.

First consider
\eqn\ans{\eqalign{W_\mu&=W_\mu(x,X(t))\cr
\chi&=\delta_aW_\mu\Gamma^\mu\eps_+\lambda^a(t)\cr}}
with $\lambda^a$ satsifying
\eqn\fccid{-i\lambda^a{\CJ_a}^b=\lambda^b}
due to \fzmid.
Denoting  $n_\p$ as the number of time derivatives and $n_f$ as the
number of fermions, the collective co-ordinate expansion is an expansion
in $n=n_\p+{1\over 2}n_f$.
The effective action obtained by substituting the ansatz
\ans\ into the action is order $n=2$. To have a consistent expansion
we must ensure that the ansatz solves the equations of motion to
order $n=0, {1\over 2},1$. To order $n=0,{1\over 2}$
\ans\ solves the equations of motion
trivially.
Just as in the bosonic case \anso, to ensure the
equations of motion are solved to order $n=1$,
it is necessary to supplement the naiive ansatz \ans\
with
\eqn\anst{\eqalign{
A_0&=\dot X^a\eps_a-i\phi_{ab}\lambda^{\dagger a}\lambda^b\cr
P&=i\phi_{ab}\lambda^{\dagger a}\lambda^b\cr}}
where $\eps_a$ and $\phi_{ab}$ were defined in the last section.
The form of these terms is motivated by considering \eom\ and \curvt.
Since we have been working in the $A_0=0$ gauge the ansatz \ans-\anst\
should really be gauged transformed into this gauge. However such an
ansatz does not seem to be as simple to write down.

After substituting the ansatz into the action \symt, integrating
over the spatial degrees of freedom and using the various results
of the last section we obtain
\eqn\seff{S_{\rm eff}={1\over 2}\int dt \CG_{ab}\{\dot X^a \dot X^b
+4i\lambda^{\dagger a}D_t\lambda^b\}-4\pi k}
where the covariant derivative is defined as
\eqn\covy{D_t \lambda^b=\dot\lambda^b+\Gamma^b_{ac}\dot X^a \lambda^c}
and the Christoffel connection is given in \chris. The constant term
$4\pi k$ is simply the energy of the static $k$ monopole configuration.
Using complex co-ordinates (based on $\CJ$) it is convenient to redefine
a real set of independent fermions $\psi^a$
via
\eqn\si{\psi^\alpha={\sqrt 2}\lambda^\alpha,\qquad\psi^{\bar\alpha}=
{\sqrt 2}\lambda^{\dagger\bar\alpha}}
(note that because of \fccid\
$\lambda^{\bar\alpha}=\lambda^{\dagger\alpha}=0$).
The effective action can then
be written in the form
\eqn\sefft{S_{\rm eff}={1\over 2}\int dt \CG_{ab}\{\dot X^a \dot X^b
+i\psi^{a}D_t\psi^b\}-4\pi k}

Since the metric on $\CM_k$ is hyperK\"ahler, the effective action
is invariant under $N=4$ worldline supersymmetry:
\eqn\sustc{\eqalign{
\delta X^a&=i\beta_4\psi^a+i\beta_m\psi^b{\CJ^{(m)}_b}^a\cr
\delta\psi^a&=-\dot X^a\beta_4+\beta_m\dot X^b {\CJ^{(m)}_b}^a\cr
}}
where $\beta_m,\beta_4$ are four real Grassmann odd worldline parameters
and $\CJ^{(m)}$ are the three complex structures on $\CM_k$ introduced
in \hkt. Not surprisingly, the origin of these supersymmetries is
precisely the unbroken supersymmetries of the field theory. The unbroken
supersymmetries are parametrised by a four dimensional chiral spinor
$\alpha_+$ which has four independent real components.
Writing
\eqn\pig{\eqalign{\delta W_\mu&=\delta X^a\delta_a W_\mu\cr
\delta\chi&=\delta_a W_\mu\Gamma^\mu\eps_+\delta\lambda^a+
s_a(\delta W_\mu)\Gamma^\mu\eps_+\lambda^a\cr}}
we substitute into the left hand side a supersymmetry transformation
\susyt\ with
parameter $\sigma\eps_+$ or $\rho C\eps_+^*$,
where $\sigma$ and $\rho$ are complex one component
Grassmann odd objects and the matrix $C$ was introduced after \sbzm.
After some lengthy algebra one obtains (to order $n=1$)
the transformations in \pig\ with $\beta_1,\beta_2$ and
$\beta_3,\beta_4$ related to the
real and imaginary parts of $\rho$ and $\sigma$, respectively.

In conclusion the effective action governing the low-energy
dynamics of the monopoles of N=2 supersymmetric Yang-Mills is
given by an N=4 supersymmetric quantum mechanics based on
the moduli space of static BPS monopoles.

\newsec{Quantisation of the effective action}

In the collective co-ordinate expansion
we neglected all of the non-zero modes.
In the quantum theory
this is equivalent to demanding
that all of these modes are in their ground states.
In particular all radiative processes are neglected.
The quantisation of the action \sefft\ describes the low energy
dynamics of $k$ supersymmetric monopoles in this approximation.
Determining the range of validity of this approximation seems a
difficult problem; we will be content to assume that it is a reasonable
approximation at low enough energies.

In \jerome\ we showed that the effective action describing
the dynamics of the lumps of the supersymmetric sigma model in $d=2+1$
is also given by \sefft. For that case the moduli space is K\"ahler
and hence the action has only $N=2$ worldline supersymmetry.
We discussed in detail in \jerome\ the quantisation of \sefft\
on a general K\"ahler moduli space and we refer the reader to that paper
for more details on the following discussion.

For the one monopole sector the moduli space
is given by $R^3\times S^1$
(see e.g. \gary). The $R^3$ corresponds to the location
of the monopole and the $S^1$ corresponds to the electric charge.
In the bosonic case, the quantum Hamiltonian is given by
\eqn\hamy{H={1\over 8\pi}{\bf P}^2+{1\over 8\pi}Q_E^2 +4\pi}
with ${\bf P}=-i\hbar\p/\p{\bf X}$, $Q_E=-i\hbar\p/\p\chi$ where
${\bf X}$, $0\le\chi\le 2\pi$ are co-ordinates on $R^3$ and $S^1$,
respectively.
Since $\chi$ is a periodic
co-ordinate the electric charge is quantized; one obtains
a tower of electrically charged dyon
states with charge $Q_E=n\hbar$ and mass $4\pi+ n^2\hbar^2/8\pi$.
This differs from the exact classical dyon mass formula \cole\
$\sqrt{16\pi^2+n^2\hbar^2}=\sqrt{Q_M^2+Q_E^2}$
by an amount which is assumed to be small
in the geodesic approximation (small electric charges).

We now outline the quantisation in the one monopole sector of
the supersymmetric theory i.e. the quantisation
of the effective action
\sefft\ on the target $R^3\times S^1$. Using complex co-ordinates,
the fermionic
commutation relations
$\{\psi^\alpha,\psi^{\bar\beta}\}=\hbar\delta^{\alpha{\bar\beta}}$
lead to a Hilbert space consisting of four types of states:
\eqn\fermc{f_1(X)|0>\qquad f_{\bar\alpha}\psi^{\bar\alpha}|0>\qquad
f_4(X)\psi^{\bar 1}\psi^{\bar 2}|0>}
where $\psi^\alpha|0>\equiv 0$. The Hamiltonian acting on these
states is given by \hamy. Thus each dyon state in the bosonic theory
is now part of a degenerate dyon multiplet. The spins of these
states can be calculated by constructing the semi-classical
spin operator along the lines of \osb. It was noted in \osb\
that the multiplet consists of
monopoles of spin ${1\over 2}$ and two with with spin $0$.
In fact this multiplet provides a representation of the
supersymmetry algebra \susyalg\ with the bound \bbb\ saturated.
Since the classical masses of all the
particles in the spectrum (monopoles, dyons, elementary particles)
saturate the bound and since the
representations of \susyalg\ are larger when the
bound is not saturated,
Witten and Olive conjectured \witol\ that there
are no quantum corrections to the classical masses.

The results of the previous section allow us to address the
low-energy dynamics of multi monopole or dyon configurations.
To do this we need to quantise the effective
$N=4$ supersymmetric quantum mechanics action \sefft\ on
the moduli space of $k$-monopoles, $\CM_k$.
In \jerome\ we discussed in detail the quantisation
of \sefft\ on a general K\"ahler moduli space.
It was noted there that for general
K\"ahler manifolds there is an operator ordering
ambiguity in the quantisation procedure; the Hilbert space of
states is either
anti-holomorphic forms or spinors on the moduli space. In the present
case where the target manifold is hyperK\"ahler and hence Ricci
flat, these two quantisations are equivalent and so there is no
ambiguity.

For definiteness we will work with anti-holomorphic forms,
although for more detailed calculations it may be more
convenient to use spinors.
The Hilbert space of states is thus given by
(superpostions of) anti-holomorphic differential forms
on $\CM_k$:
\eqn\hilb{\eqalign{
|f>&=\psi^{\bar\alpha_1}\dots\psi^{\bar\alpha_p}|0>{1\over p!}
f_{\bar \alpha_1\dots \bar \alpha_p}\cr
|f> &\leftrightarrow {1\over p!}
f_{\bar \alpha_1\dots \bar \alpha_p}dZ^{\bar \alpha_1}
\wedge\dots\wedge dZ^{\bar \alpha_p}\cr
}}
where $\{Z^{\alpha},Z^{\bar \alpha}\}$ are complex co-ordinates on $\CM_k$
and $\psi^\alpha|0>=0$.
The degree of the differential form
corresponds to the number of fermionic zero modes that are excited.
The quantum hamiltonian is given by the Laplacian acting on differential
forms
\eqn\hamilt{
H=\hbar^2({\bar\p}^\dagger{\bar\p}+{\bar\p}{\bar\p}^\dagger)+4\pi k=
{\hbar^2\over 2}(dd^\dagger+d^\dagger d)+4\pi k}
where $d$ is the exterior derivative, $d^\dagger$ is its adjoint,
$\bar\p$ and $\bar \p^\dagger$ are the anti-holomorphic analogues and
the equality uses the K\"ahler property of the moduli space.

Asymptotically, $\CM_k$ is isomorphic to
$k$ copies of $\CM_1$ thus providing the
interpretation of $k$ well separated monopoles. In this region an
anti-holomorphic form with non-zero asymptotic support
can be written as (a sum of) wedge products of
forms on each of the $k$ copies of $\CM_1$:
\eqn\asympt{|f>\leftrightarrow f\approx f^{(1)}\wedge\dots\wedge f^{(k)}
}
where $f^{(i)}$
is a differential form on $\CM_1$ (a state of the form \fermc).
The interpretation of these
states is that of well separated monopoles with fermionic zero
modes excited on each. They can thus be identified with (superpositions of)
monopoles and dyons in the spectrum.

To proceed further we use the fact \hitch\ that
$\CM_k$ isometrically decomposes as
\eqn\iso{\CM_k=R^3\times\left({S^1\times\CM_k^0\over Z_k}\right).}
The $R^3$
corresponds to the location of the centre of mass and the
$S^1$ corresponds to the total electric charge of the $k$
monopoles. The hyperK\"ahler manifold $\CM_k^0$ is the
interesting part and determines the relative motion. The presence of the
$Z_k$ is a reflection of the fact that the monopoles are not classically
distinguishable. First consider quantising
on $\CM_1\times \CM_k^0$.
Since $\CM_k^0$ is hyperK\"ahler,
the quantum states can be written as the wedge product
of states of the form \fermc\ with antiholomorphic
differential forms on $\CM_k^0$.
The quantum Hamiltonian is given
by $H=H_{\rm COM}+H^0$ where $H_{\rm COM}$ is the free centre of
mass Hamiltonian, up to some factors of the form \hamy\ (no factor
of $4\pi$),
and $H^0$ contains the non-trivial dynamics and is
given by \hamilt\ acting on $\CM_k^0$.

For the quantisation on $\CM_1\times \CM_k^0$ to be valid
on
$\CM_k$
one needs to restrict
the states to ensure that they
are well defined differential forms on $\CM_k$:
the differential form on $\CM_k^0$ must satisfy a discrete condition
depending on the total electric charge of the state. For example, for $k=2$
we can choose co-ordinates $({\bf X},\chi,r,\theta,\phi,\psi)$ on $\CM_2$
where $(\theta,\phi,\psi)$ are Euler angles and
the action of $Z_2$ implies that we identify the point
$({\bf X},\chi,r,\theta,\phi,\psi)$ with $({\bf X},\chi+\pi,
r,\theta,\phi,\psi+\pi)$ (see \gary).
Let $\Psi$ be a state of the form \fermc\ with $f_i(X)=exp(i/\hbar(
{\bf P}{\bf X}+S\chi))$ where the integer $S$ is the total electric charge
and $\Phi$ be an antiholomorphic form on $\CM_2^0$.
For the state $\Psi\wedge\Phi$
to be well defined on $\CM_2$ the antiholomorphic form on $\CM_2^0$ must
satisfy
$\Phi(r,\theta,\phi,\psi+\pi)=(-1)^S\Phi(r,\theta,\phi,\psi)$.
Such a restriction on the forms on $\CM_k^0$ is to be implicitly understood
in the following discusion.

The scattering theory can be investigated by constructing
non-normalisable eigenforms of the Laplacian on $\CM_k^0$. In the asymptotic
region a decomposition of the form \asympt\ would allow the
identification of these states with well separated monopoles and
dyons in the spectrum. For $\CM_2^0$, the only case where the
moduli space has been explicitly constructed,
this seems a difficult undertaking.

The bound states are determined by the normalisable eigenforms
on $\CM_k^0$. Since these states are tensored with a state
of the form \fermc, this means that any
bound state is actually part of a supermultiplet of bound states.
Bound states of the bosonic theory, discussed in \gary,
automatically become part
of a multiplet of bound states in the supersymmetric theory,
corresponding to zero forms and $4k-4$ forms on $\CM_k^0$.
In general there will be a whole slew of bound states. It seems likely
that the lowest energy bound state, for a given total electric charge,
will be a true bound state of the
full quantum field theory, since there isn't anything it
could decay into.

We can be more explicit about a particularly important class
of bound states. Fixing the total electric charge the energy of
a bound state is bounded below by $4\pi k$ plus the energy coming
from the centre of mass contribution. For a state to attain this
bound, from
\hamilt\ we see that the differential form on $\CM_k^0$
must be annihilated by both the exterior derivative $d$ and
its adjoint $d^\dagger$.
Such anti-holomorphic forms are equivalent
to the dolbeault cohomolgy classes of $\CM_k^0$.
Since there are no other bound states with lower energy it is
plausible that these normalisable cohomology classes\foot{If we had
described the Hilbert space using spinors on moduli space, these bound states
would be given by zero modes of the Dirac operator on $\CM_k^0$.}
correspond to bound states in the full
quantum field theory. This interpretation would be strengthened
if we could show that the energy of the states is strictly
less than the continuum (as opposed to being the lowest energy state
bounding the continuum).
Unfortunately it
is not known to the author whether such normalisable harmonic forms
exist on $\CM_k^0$; perhaps it is not so difficult to
determine if they exist
on the Atiyah-Hitchin space $\CM_2^0$.

\newsec{Discussion}
We have shown that the dynamics of $k$ monopoles of $N=2$ supersymmetric
Yang-Mills theory are determined by an effective
$N=4$ supersymmetric quantum
mechanics based on the moduli space of $k$ monopoles. The four worldline
supersymmetries come from the fact that half of the supersymmetries
in the field
theory (described by four real parameters)
are unbroken by a monopole configuration.
We discussed how the Hilbert space of states of the effective theory are
given by holomorphic differential forms on the moduli space and how
the normalisable dolbeault cohomology
classes of $\CM^0_k$
are related to bound states of the full quantum field theory.
It would be interesting to investigate the existence of such
classes. The scattering problem requires the construction of
eigenforms of the Laplacian on moduli space.
Since the metric on moduli space is explicitly
known for
the case of two monopoles, perhaps some further progress could be made
on these topics.

It would also be interesting to generalise the results of this
paper to study the dynamics of the monopoles of $N=4$ supersymmetric
Yang-Mills. Although there appear to be some technical complications
in repeating the steps in this paper the result seems almost certain.
A monopole configuration will still only break half
of the supersymmetries of the field theory.
Since there are now twice as many supersymmetries the unbroken
supersymmetries
are determined by eight real parameters instead of four.
We thus expect the low energy dynamics to be described by an
$N=8$ supersymmetric quantum mechanics
(eight real parameters)
based on the moduli space $\CM_k$. In addition the Callias index
theorem says there will be twice as many fermionic zero modes
as in the $N=2$ model.
Putting this together we are led to an effective action of the
form
\eqn\sefffour{S={1\over 2}\int dt \left[\dot X^a \dot X^b {\CG}_{ab}
+i\bar\psi^{a}\gamma^0D_t\psi^b{\CG}_{ab}+{1\over 6}{\CR}_{abcd}(\bar\psi^{a}
\psi^c)
(\bar\psi^{b}
\psi^d)\right]}
where $\psi^a$ is a two component real spinor
(i.e. two worldline Grassmann numbers), $\gamma^0=\sigma^2$
and ${\CR}_{abcd}$ is the Riemman tensor of $\CG$. This
action is the dimensional reduction of
the two dimensional supersymmetric sigma model and it
is well known that when the metric
is hyperK\"ahler it admits four supersymmetries, each parametrised
by a two component Majorana spinor \alvy.
Thus \sefffour\ based on $\CM_k$ does
indeed admit eight worldline supersymmetries (it is a matter of convention
to describe this action as having $N=4$ or $N=8$ worldline
supersymmetry; we stick to the nomenclature that $N$ equals the
number of real worldline supersymmetry parameters).
The quantisation of this model leads
to a Hilbert space of states consisting of real differential forms
on the moduli space \weet. In this case the real normalisable
harmonic forms of $\CM^0_k$ will be
related to bound
states of the full quantum field theory.

An interesting feature of the $N=4$ model is that the monopole
spectrum contains particles of spin 1 \osb. Consequently, one
might expect the
amplitudes for the scattering of such monopoles to obey Ward identities.
It would be very interesting if such identities were encoded in the
geometry of moduli space. Another reason for pursuing more detailed
calculations for this model
is that it may be possible
to gain some insight into the
Montonnen and Olive duality conjecture
\montolive,\osb.
In particular this conjecture says that
low energy scattering of monopoles at weak coupling should correspond
to low energy scattering of elementary particles at strong coupling.
We can possibly make some progress on the monopole scattering
but a direct comparison with the elementary particle
scattering is not possible because it is at strong coupling.
Nevertheless, the discovery of Ward identities mentioned above
would possibly be further evidence that the conjecture is correct.

\bigskip\centerline{\bf Acknowledgements}\nobreak
It is a pleasure to thank Jeff Harvey for many useful discussions.
In addition conversations with Gary Gibbons, Nick Manton, Trevor
Samols and Berndt Schroers are gratefully acknowledged.
This work is supported by a grant from the Mathematical Discipline Center
of the Department of Mathematics, University of Chicago.
\listrefs
\end